\documentclass[aps,prl,twocolumn,nofootinbib,amsfonts,pdftex,floatfix,superscriptaddress]{revtex4-1}
\usepackage{xcolor}
\usepackage{amsmath}
\usepackage{graphicx}
\usepackage{amsmath}
\usepackage{hyperref}
\hypersetup{
    colorlinks,
    linkcolor={red!50!black},
    citecolor={blue!50!black},
    urlcolor={blue!80!black}
}

\usepackage[utf8]{inputenc}
\usepackage[english]{babel}

\date{\today}

\begin{document}

\title{\texorpdfstring{$\mathbf{\beta}$}{beta}-delayed proton emission from \texorpdfstring{$\mathbf{^{11}}$}{11-}Be in 
effective field theory}

\author{Wael Elkamhawy}
\email{elkamhawy@theorie.ikp.physik.tu-darmstadt.de}
\affiliation{Institut f\"ur Kernphysik, Technische Universit\"at Darmstadt, 64289 Darmstadt, Germany}

\author{Zichao Yang}
\email{zyang32@vols.utk.edu}
\affiliation{Department of Physics and Astronomy, University of Tennessee, Knoxville, TN 37996, USA}

\author{Hans-Werner Hammer}
\email{Hans-Werner.Hammer@physik.tu-darmstadt.de}
\affiliation{Institut f\"ur Kernphysik, Technische Universit\"at Darmstadt, 64289 Darmstadt, Germany}
\affiliation{ExtreMe Matter Institute EMMI, GSI Helmholtzzentrum
für Schwerionenforschung GmbH, 64291 Darmstadt, Germany}

\author{Lucas Platter}
\email{lplatter@utk.edu}
\affiliation{Department of Physics and Astronomy, University of Tennessee, Knoxville, TN 37996, USA}
\affiliation{Physics Division, Oak Ridge National Laboratory, Oak Ridge, TN 37831, USA}

\date{\today}

\begin{abstract}
  We calculate the rate of the rare decay $^{11}$Be into
  $^{10}\text{Be} + p +e^- + \bar{\nu}_e$ using Halo effective field
  theory, thereby describing the process of beta-delayed proton
  emission. We assume a shallow $1/2 ^+$ resonance in the
  $^{10}$Be$-p$ system with an energy consistent with a recent experiment
  by Ayyad {\it et al.} and obtain
  $b_p = 4.9_{-2.9}^{+5.6}\text{(exp.)}_{-0.8}^{+4.0}\text{(theo.)}
  \times 10^{-6}$ for the branching ratio of this decay, predicting a 
  resonance width of $\Gamma_R = (9.0^{+4.8}_{-3.3}\text{(exp.)}^{+5.3}_{-2.2}\text{(theo.)})$~keV.
  Our calculation shows that the experimental branching ratio and 
  resonance parameters of Ayyad {\it et al.} are consistent with each other.
  Moreover, we analyze the general impact of a resonance on the
  branching ratio and 
  demonstrate that a wide range of combinations of resonance energies and 
  widths can reproduce branching ratios of the correct order. 
  Thus, no exotic mechanism (such as beyond the standard model physics) is 
  needed to explain the experimental decay rate.
\end{abstract}

\maketitle

\paragraph{\bf Introduction.}
Halo nuclei display a large separation of scales between a few loosely
bound halo nucleons and a tightly bound
core~\cite{Hansen:1995pu,Jonson:2004,Riisager:2012it,Tanihata:2016zgp}.
The emergence of the halo degrees of freedom is a fascinating aspect
of nuclei away from the valley of stability. 
The halo nucleons in the core potential spend most
of their time in the classically forbidden region outside of the range
of the core potential. This is analog to the tunnel effect.
But since the halo nucleons are bound to the core, they always
have to come back into the core potential.
This separation of
scales can be used to treat these systems using an effective field
theory (EFT) approach called Halo EFT
\cite{Bertulani:2002sz,Bedaque:2003wa,Hammer:2017tjm}.  Common to all
EFTs is that observables are described in a systematic low-energy
expansion and that the accuracy of a calculation can be systematically
improved. Halo EFT has been applied to a number of observables,
including electromagnetic capture reactions and photodissociation
processes
\cite{Hammer:2011ye,Rupak:2011nk,Ryberg:2014exa,Zhang:2015ajn,Higa:2016igc,
  Premarathna:2019tup,Zhang:2019odg}.

Here we will consider, for the first time, the weak decay of the
valence neutron of the halo nucleus $^{11}$Be into the continuum,
${^{11}\text{Be}} \rightarrow {^{10}\text{Be}}+ p + e^{-}
  + \bar{\nu}_{e}$,
within Halo EFT.

First experimental results for this rare decay mode were presented in
Refs.~\cite{Borge:2012nz,Riisager:2014rma}.  Riisager et
al.~\cite{Riisager:2014gia} measured a surprisingly large branching
ratio for this decay process, $b_p = 8.3(9)\times 10^{-6}$, which
could only be understood in their Woods-Saxon model analysis if the
decay proceeds through a new single-particle resonance in $^{11}$B.
Their measured branching ratio is also more than two orders of magnitude larger
than the cluster model prediction by Baye and
Tursunov~\cite{Baye:2010cj}.  This led Pf\"utzner and Riisager
\cite{Pfutzner:2018ieu} to suggest that $\beta$-delayed proton
emission in $^{11}$Be is also a possible pathway to detect a dark
matter decay mode as proposed by Fornal and Grinstein
\cite{Fornal:2018eol}. More recently, this branching ratio was
remeasured by Ayyad {\it et al.}~\cite{Ayyad:2019kna} as
$b_p = 1.3(3)\times 10^{-5}$, similar in size to the previous
measurement. They also presented new evidence for a low-lying
resonance in $^{11}$B with resonance energy $E_R = 0.196(20)$~MeV and
width $\Gamma_R = 12(5)$~keV. Using these parameters, the authors
calculated the decay rate in a Woods-Saxon model assuming a pure
Gamow-Teller transition. They obtained $b_p = 8 \times 10^{-6}$, which
has the correct order of magnitude but is only consistent within a
factor of two with their experimental result. The work by Ayyad et
al. was criticized in a recent comment by Fynbo et
al.~\cite{Fynbo:2019dfo}.  A new experiment by Riisager et
al.~\cite{Riisager:2020glj} gives an upper limit of
$b_p \leq 2.2 \times 10^{-6}$ for the branching ratio but some
questions remain due to inconsistencies between different
measurements. In conclusion, the branching ratio for $\beta$-delayed
proton emission in $^{11}$Be remains an important unsolved problem.

The ground state of $^{11}$Be is a well-understood $S$-wave halo
nucleus. From the ratio of the one-neutron separation energy of
$^{11}$Be and the excitation energy of the $^{10}$Be core, one can
extract the expansion parameter for a description with the core and
valence neutron as effective degrees of freedom,
$R_{\text{core}}/R_{\text{halo}}\approx 0.4$~\cite{Hammer:2011ye}.
Here $R_{\text{core}}$ and $R_{\text{halo}}$ are the length scales of
the core and halo, respectively.  In principle, both the $^{10}$Be
core and the halo neutron can $\beta$-decay. Since the half-life of
the neutron ($T_{1/2}=10$~min) is much shorter than the half-life of
the core ($T_{1/2}=10^6$~a), it is safe to assume that for
$\beta$-delayed proton emission it is always the halo neutron that
decays in the halo picture.  Therefore, one would naively expect the
nucleus to emit this proton due to the repulsive Coulomb interaction:
  ${^{11}\text{Be}} \rightarrow {^{10}\text{Be}}+ p + e^{-}
  + \bar{\nu}_{e}$~.
This process, called $\beta$-delayed proton emission, has
well-defined experimental signatures. However, it is also known
that short-distance mechanisms such as the decay into excited states
of ${}^{11}$B (that are beyond the halo interpretation)
dominate the total $\beta$-decay rate of $^{11}$Be
\cite{Refsgaard:2018yaa, Kelley:2012qua}.

Halo EFT offers
a new perspective on $\beta$-delayed proton emission
from $^{11}$Be by providing a value for the decay rate with
a robust uncertainty estimate.
It uses the appropriate degrees of freedom
and parametrizes the decay observables in terms of a few
measurable parameters.
Thus, it is perfectly suited for the theoretical
description of low-energy processes such as 
$\beta$-delayed proton emission from halo nuclei.
Kong and Ravndal \cite{Kong:1999tw}
used
these ideas to successfully describe the inverse
process of $pp$-fusion into a deuteron and leptons. 
In contrast to the previous calculation
  in Ref.~\cite{Baye:2010cj},
we will use new
experimental input parameters and put additional emphasis on
the uncertainties associated with using effective degrees of
freedom. The halo neutron can $\beta$-decay through both the
  Gamow-Teller and Fermi operators. The Fermi operator can only connect
  states in the same isospin multiplet. If all neutrons in $^{11}$Be
  contribute to the $\beta$-decay, this implies that the final state
  must have $T=3/2$ for a Fermi transition.
  No such states are currently known in $^{11}$B within the
  $\beta$-decay window. However, due to the halo character of $^{11}$Be
  we expect that only the halo neutron decays, such that the final
  state has no definite isospin. Thus, we will keep our analysis general and
  consider both the scenarios of Gamow-Teller and Fermi decay as well 
a pure Gamow-Teller decay in the following.
  Specifically, we will show that based on the measured branching
ratio, a low-lying resonance is the likely reason for the large
partial decay rate, confirming the suggestion of
  Ref.~\cite{Riisager:2014gia}. 
Furthermore, in $^{11}$B, we explore the impact of
the resonance energy and width on the decay rate and show that the
recent results for the resonance energy and width of a low-lying resonance are
consistent with the experimentally measured branching ratio.

In order to keep our presentation self-contained,
we start by summarizing the concepts of Halo EFT for $S$-wave halo
nuclei. 
We discuss the calculation of decay rates with and without
resonant final state interactions and then display our results. Note that these are EFTs for two different scenarios. Formally, we perform calculations up to corrections of order $R_{\text{core}}/R_{\text{halo}}$ in both scenarios but because of the different physics assumptions these cannot be directly compared. We
conclude with a summary.

\paragraph{\bf Theoretical foundations.}
The Halo EFT Lagrangian $ \mathcal{L}$ for $^{11}$Be as well as the 
low-lying resonance in $^{11}$B up to next-to-leading
order can be
written as $\mathcal{L} = \mathcal{L}_0 + \mathcal{L}_{d}$,
where $\mathcal{L}_0$ is the free Lagrangian of the $^{10}$Be core,
neutron and proton
\begin{equation}
\begin{aligned}
  \label{freeLagrangian}
  \mathcal{L}_0 = c^\dagger&\left(i\partial_t + \frac{\nabla^2}{2m_c}
  \right)c
  +   n^\dagger\left(i\partial_t + \frac{\nabla^2}{2m_n} \right)n \\
  &+   p^\dagger\left(i\partial_t + \frac{\nabla^2}{2m_p} \right)p ~,
\end{aligned}
\end{equation}
with $c$, $n$ and $p$ the core, neutron and proton fields, respectively. The 
masses of core, neutron and proton are denoted by 
$m_c=9327.548$~MeV, $m_n=939.565$~MeV and $m_p=938.272$~MeV.
The $S$-wave core-neutron as well as core-proton interaction
are described by  $\mathcal{L}_{d}$, which reads
\begin{equation}
\begin{aligned}
  \label{eq:Lsigma}
  \mathcal{L}_{d}= d_{\text{Be}}^\dagger&\left[ \eta \left( i\partial_t +
      \frac{\nabla^2}{2M_{nc} } \right)+ \Delta \right] d_{\text{Be}} \\
      &+d_{\text{B}}^\dagger\left[ \tilde{\eta} \left( i\partial_t +
      \frac{\nabla^2}{2M_{pc} } \right)+ \tilde{\Delta} \right] d_{\text{B}} \\
  &-g  \left[ c^\dagger n^\dagger d_{\text{Be}} + \text{H.c.}  \right]
  -\tilde{g}  \left[ c^\dagger p^\dagger d_{\text{B}} + \text{H.c.}  \right],
\end{aligned}
\end{equation}
where $d_{\text{Be}}$ and $d_{\text{B}}$ are dimer fields, with spin indices suppressed, that
represent the $J^P = 1/2^{+}$ ground state of $^{11}$Be 
and the $J^P = 1/2^{+}$ low-lying resonance in $^{11}$B, respectively, while
$M_{nc}=m_n+m_c$ and $M_{pc}=m_p+m_c$.

The renormalization of the low-energy constants for $^{11}$Be has 
been discussed in Ref.~\cite{Hammer:2011ye}. Here, we will briefly summarize
the relevant results to define our notation.
Due to the non-perturbative nature of the interaction, we need to
resum the self-energy diagrams to all orders. After matching the
low-energy constants for $^{11}$Be appearing in Eq.~\eqref{eq:Lsigma} 
to the effective range expansion, we obtain the full two-body $T$-matrix 
\begin{equation}
  \label{eq:t0}
  T_0(E)  
  =\frac{2\pi}{m_R} \left[\frac{1}{a_0} - r_0 m_R E - \sqrt{-2m_R E -i \epsilon} 
\right]^{-1}~.
\end{equation}
where 
$m_R$ is the reduced mass, and 
$a_0$, $r_0$ are the $S$-wave $^{10}$Be$-n$ scattering length and effective 
range, respectively.
The
residue at the bound state pole of Eq.~(\ref{eq:t0}) is 
required to calculate physical observables,
$Z=\textstyle{\frac{2\pi \gamma_0}{m_R^2}}/(1-r_0\gamma_0)$~,
with $\gamma_0 =(1-\sqrt{1-2r_0/a_0})/r_0\equiv
\sqrt{2m_R S_n}$ the binding momentum of the $S$-wave
halo state, and $S_n$ the one-neutron separation energy of the halo nucleus.

In order to investigate $\beta$-delayed proton emission from
$^{11}$Be, we include the weak interaction current allowing
transitions of a neutron into a proton, electron and antineutrino which
corresponds to the hadronic one-body current.
Moreover, we have to consider hadronic two-body currents that appear in the 
dimer formalism once the effective range is included.
The corresponding Lagrangian is given by
\begin{align}
  \mathcal{L}_{\text{weak}} = -\frac{G_F}{\sqrt{2}} l_-^\mu \left(\left(J_\mu^+\right)^{\text{1b}} + \left(J_\mu^+\right)^{\text{2b}}\right)~,
\end{align}
where $l^{\mu}_- = \bar{u}_e \gamma^\mu (1-\gamma^5) v_{\bar{\nu}}$ 
and $\left(J_{\mu}^{+}\right)^{\text{1b}} = (V_\mu^1 - A_\mu^1)  + i (V_\mu^2 - A_\mu^2) $ 
denote the leptonic and hadronic one-body currents, respectively. 
Here the hadronic one-body current is decomposed into vector and axial-vector
contributions. At leading order, the contributions to this current are 
$ V_0^a  = N^\dagger \frac{\tau^a}{2}N$, 
$A_k^a = g_A N^\dagger \frac{\tau^a}{2}\sigma_k N$, 
where $|g_A| \simeq 1.27$ is the ratio of the axial-vector to vector
coupling constants~\cite{PhysRevD.98.030001}. Terms with more derivatives and/or more fields
(many-body currents) will appear at higher orders. The first and
second term give the conventional Fermi and Gamow-Teller operators,
respectively.
Including resonant core-proton final state interactions, we have to take into account a two-body current with known coupling constants which arises from gauging the time derivative of the dimer fields appearing in Eq.~\eqref{eq:Lsigma}.
It is also decomposed into vector and axial-vector contributions and reads
\begin{align}
	\left(J_\mu^+\right)^{\text{2b}} = \begin{cases} -d_{\text{B}}^\dagger \, d_{\text{Be}} \,\, &\mu=0~,\\ g_A \, 	d_{\text{B}}^\dagger \, \sigma_k \, d_{\text{Be}} \,\, &\mu=k=1,2,3~.\end{cases}
	\label{eq:dimerterm}
\end{align}
In addition, there is also an unknown contribution usually denoted as $L_{1A}$ that normally appears at the same order.
However, in the case with Coulomb interaction, this piece is suppressed by $\left(R_{\text{core}}/R_{\text{halo}}\right)^{1/2}$ compared to the two-body current in Eq.~\eqref{eq:dimerterm}.\footnote{The scaling of $r_0^C \sim 1/k_C$ leads to the suppression of the counterterm contribution $L_{1A}$.} Therefore, it contributes only at NNLO allowing us to make predictions up to NLO. Note that our power counting including resonant final state interactions implies a suppression of $\left(R_{\text{core}}/R_{\text{halo}}\right)^{1/2}$ going from order to order instead of $R_{\text{core}}/R_{\text{halo}}$ as in the case without resonant final state interactions.
\paragraph{\bf Weak matrix element and decay rate.}
We ignore recoil effects in the $\beta$-decay and take
both the Gamow-Teller and Fermi transitions into account. After
lepton sums, spin averaging, and partial phase space integration,
we obtain the decay rate
\begin{multline} \label{eq:decayRate}
\Gamma =\frac{G_F^2 (1+3g_A^2)}{4 \pi^5} \int dp \int dp_e  p^2 p_e^2(E_0 -E - E_e)^2   \\
\times %F_e(Z,E_e)
 C^2(\eta_e)~ \overline{|\mathcal{A}(\mathbf{p})|^2} ~ 
\Theta(E_0 - E -E_e)~,
\end{multline}
where $\mathcal{A}$ is the {\it reduced} hadronic amplitude for
Gamow-Teller and Fermi transitions whose operator coefficients have
been factored out and $\Theta$ is the Heaviside step function. 
Moreover, ${\bf p}$ is the relative momentum
of the outgoing proton and core, while $E = p^2/(2m_R)$ is their kinetic
energy. Furthermore,
$E_0 = \Delta m - S_n$, where $\Delta m = 1.29$ MeV is the mass
difference between neutron and proton,
and $E_e = \sqrt{m_e^2 + p_e^2}$ is the energy of the
electron with $m_e=0.511$~MeV denoting the electron mass.

\begin{figure}[t]
  \centerline{\includegraphics[width = 0.49 \textwidth]{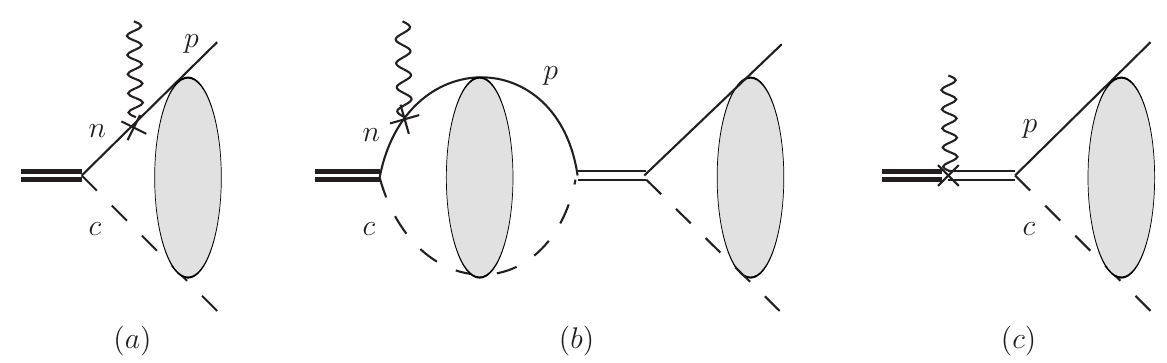} }
    \caption{\label{fig:Feynman} $(a)$: Feynman diagram for
      the weak decay of a one-neutron halo nucleus
      into the corresponding core and a proton with Coulomb final
      state interactions only. $(b)+(c)$: Contributions of
      resonant final state interactions. The thin double
      line in the middle denotes the dressed $^{10}$Be$-p$
      propagator. The shaded ellipse denotes the Coulomb Green's function.}
\end{figure}

The Sommerfeld factor of the electron is given by
\begin{align}
  \label{eq:Sommerf}
C^2(\eta_e)= \frac{2\pi\eta_e}{(e^{2\pi\eta_e} - 1)}~,
  %	F_e(Z,E_e) = \frac{2\pi\eta_e}{(e^{2\pi\eta_e} - 1)}~,
\end{align}
where $\eta_e = \alpha Z Z_e E_e/|\mathbf{p_e}|$
%is the Sommerfeld parameter of the electron
with $\alpha =1/137$ the fine structure 
constant. We use $Z=Z_p$ in order to ensure that we reproduce the 
free neutron decay width in the limit of a vanishing one-neutron 
separation energy of $^{11}$Be. This means that the electron is only
interacting with the outgoing proton. For the leading 
contributions resulting from diagrams (a) and (b) of Fig.~\ref{fig:Feynman},
we assume this to be a good approximation since the $^{10}$Be core is far 
away from the decaying valence neutron due to the small one-neutron
separation energy. For consistency, we use the same Sommerfeld factor
in diagram (c) of Fig.~\ref{fig:Feynman} although it contains interactions
of all particles at a single space-time point. However, since diagram (c) is
subleading, we also expect a deviation of subleading order.
In order to confirm this expectation,
we have performed an explicit calculation  using $Z=Z_p+Z_c$ for all
diagrams, which leads to a change  of order 30\% in the decay rate.
Thus, this effect is beyond the 40\% accuracy of our calculation
(see below) and would enter at higher orders.
If a pure Gamow-Teller transition is considered, the
  factor $1+3g_A^2$ is replaced by $3g_A^2$. This results in a reduction
  of the decay rate by 17~\%.

\paragraph{\bf Beta-strength sum rule.} The so-called Fermi and
Gamow-Teller sum rules (also collectively known as beta-strength sum
rule) count the number of weak charges that can decay in the initial
state. We will require that this beta-strength sum rule is fulfilled exactly at
each order within our EFT power counting. 
The beta-strenghts are related to the comparative half-life
of a decay, the so-called $ft$ value given by
\begin{align}
	ft = \frac{B}{B_{\text{F}}+g_A^2 B_{\text{GT}}}\,,
\end{align}
where $B=2\pi^3\ln2/(m_e^5G_F^2)$ is the $\beta$-decay constant. In this paper, we will use the value $B=6144.2$~s~\cite{RevModPhys.84.567,Hardy:2004id}.
With $B_{\text{GT}}=3B_{\text{F}}$, we find
\begin{align}
  B_{\text{F}}=\frac{B}{(1+3g_A^2)}\frac{1}{ft}\,.
\end{align}
The inverse $ft$ value is directly related to the transition matrix element $\mathcal{M}$ of $^{11}$Be into $^{10}\text{Be}+p$,
\begin{equation}
\begin{aligned}
  \frac{1}{ft}&=\frac{1}{B} ~ \overline{\left|\mathcal{M}\right|^2}\\
  &=\frac{1}{B} \frac{(1+3g_A^2)}{2\pi^2} \int dE \,  m_R \sqrt{2 m_R E} ~ \overline{|\mathcal{A}(\mathbf{p})|^2}\,.
\end{aligned}
\end{equation}
For a transition into the
continuum, the sum rule is exactly fulfilled when integrating the
differential beta-strengths
\begin{align}
	\frac{dB_{\text{F}}}{dE} &= \frac{1}{2\pi^2} ~ m_R \sqrt{2 m_R E} ~~ 
	\overline{|\mathcal{A}(\mathbf{p})|^2} \,,\\
	\frac{dB_{\text{GT}}}{dE} &= 3 \frac{dB_{\text{F}}}{dE}~,
\end{align}
over the whole continuum leading to the sum rules \mbox{$B_{\text{F}}=1$} and $B_{\text{GT}}=3$.  
In the halo picture, we therefore
expect beta-strengths $B_{\text{F}}$ and $B_{\text{GT}}$ to be at most
$1$ and $3$, respectively, when integrating over the available
$Q$-window. 
At LO where the full non-perturbative solution for a
zero-range interaction is used in the incoming as well as outgoing
channel, the sum rule is always satisfied. At NLO where range
corrections are included, the sum rule puts strong constraints on the
ranges in the incoming and outgoing channels such that only certain
combinations are allowed.

\paragraph{\bf Hadronic current without resonant final state interactions.}
The amplitude for the charge changing weak transition of a two-body
system is illustrated as diagram $(a)$ of Fig.~\ref{fig:Feynman}. It
was first calculated in pionless EFT by Kong and Ravndal
\cite{Kong:1999tw}.
%The amplitude
The corresponding hadronic current can be written as \cite{Ryberg:2013iga}
\begin{equation}
  \mathcal{A}_C^{(a)}(\mathbf{p}) =
  -i g \sqrt{Z} C(\eta_p) e^{i\sigma_0} \frac{2m_R}{\mathbf{p}^2+\gamma_0^2} 
e^{2\eta_p \arctan(|\mathbf{p}|/\gamma_0)}\,,
  \label{eq:Cb-amp}
\end{equation}
where $\sigma_0$ is the Coulomb phase and $C^2(\eta_p)$ is the
Sommerfeld factor from Eq.~(\ref{eq:Sommerf}). In the
$^{10}\text{Be}- p$ system, the Sommerfeld parameter is
$\eta_p = \alpha Z_pZ_c m_R/|\mathbf{p}|=k_C/|\mathbf{p}|$, with $Z_p = 1$ and
$Z_c = 4$.

\paragraph{\bf Hadronic current with resonant final state interactions.}
The current \eqref{eq:Cb-amp} includes only the final state
interaction from the exchange of Coulomb photons.  We now consider
resonant final state interactions whose signature is a low-lying
resonance in the $^{10}\text{Be}-p$ channel up to NLO. These
contributions are shown as diagrams $(b)$ and $(c)$ of
Fig.~\ref{fig:Feynman}. Diagram (c) contributes only at NLO
to the amplitude. It arises from a two-body current (with
known coupling strength) that appears as a result of the
energy-dependent interactions used in the initial state (see Eq.~\eqref{eq:Lsigma}) 
and the final state (see
Ref.~\cite{Higa:2008dn}). The thin double line together with the
shaded ellipses that represent Coulomb Green's functions as depicted
in diagram $(b)$ essentially combine to the strong scattering amplitude $T_{CS}$
given either in Eq.~\eqref{eq:tcs-resonance} or
\eqref{eq:tcs-neg_a}~\cite{Higa:2008dn,Kong:1999tw}.

The degrees of freedom in Halo EFT are the emitted outgoing proton and
$^{10}$Be.  Our treatment of the resonance follows
Ref.~\cite{Higa:2008dn}.  The corresponding strong scattering
amplitude modified by Coulomb corrections is~\cite{Higa:2008dn}
\begin{equation}
  \label{eq:tcs-resonance}
T_{CS} = \frac{-4\pi/m_R}{ \left(r_0^C - \frac{1 }{3k_C}\right)\left( p^2 - 
k_R^2\right) + \frac{p^2}{3k_C} - 4 k_C H(\eta_p)}~,
\end{equation}
where
$H(\eta_p)=\operatorname{Re}[\psi(1+i \eta_p)]-\ln \eta_p+\frac{i}{2 \eta_p} 
C^2(\eta_p)$~,
with the digamma function $\psi(z)$. 
The parameters in
Eq.~\eqref{eq:tcs-resonance} are directly related to the complex pole
momentum \mbox{$k^*=k_R-ik_I$}:
\begin{align}
-\frac{1}{a_0^C}& = - \left(r_0^C - \frac{1}{3k_C}\right)\frac{k_R^2}{2} ,\\
  r_0^C &=  - \frac{2\pi k_C}{k_R k_I} \frac{1}{e^{2\pi k_c/k_R} -1} + \frac{1}{3k_C} ,
\end{align}
where $a_0^C$ and $r_0^C$ are the Coulomb-modified scattering length and
effective range, respectively. Within our power counting, the parameters $k_C$, $k_R$, $k_I$ as well as $\gamma_0$ scale as $1/R_{\text{halo}}$ implying that both Coulomb-modified scattering parameters $a_0^C$ and $r_0^C$ scale as $R_{\text{halo}}$.
Note that we include $r_0^C$ despite this scaling only at
  NLO, since the range $r_0$ in the incoming channel scales as $R_{\text{core}}$.
  The inclusion of both ranges at the same order guarantees that the beta-strength sum rule is satisfied at both LO and NLO.

\begin{figure}[t]
      \includegraphics[width=0.45\textwidth]{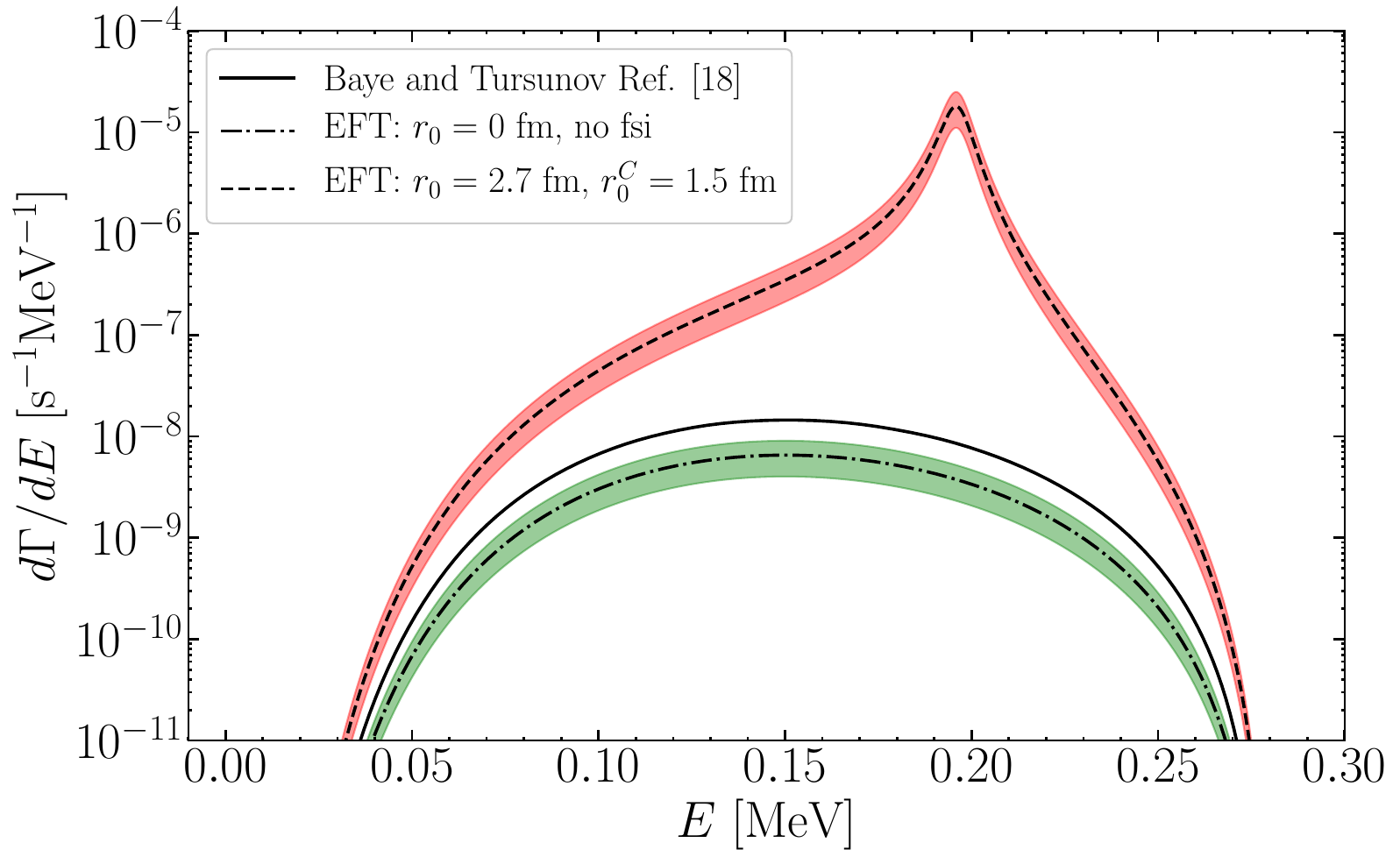}
      \caption{\label{fig:gamma-be11}Differential
        decay rate $d\Gamma/dE$ for $\beta$-delayed proton emission
        from $^{11}$Be as a
        function of the final-state particle energy $E$.
        The dash-dotted line shows our EFT result without resonant final state interactions 
        while the solid line gives the result obtained by Baye and Tursunov \cite{Baye:2010cj}.
        The dashed line shows the EFT result including a resonance
        at $E_R=0.196$~MeV in the outgoing channel at NLO. 
        The colored bands give the EFT uncertainty.
      }
\end{figure}

The diagrams $(b)$ and $(c)$ of Fig.~\ref{fig:Feynman} lead to
\begin{align}
  \label{Total}
\mathcal{A}_{CS}^{(b)}  
&= -i g\sqrt{Z} {4m_R^2} C(\eta_p) e^{i\sigma_0}  \mathcal{I} T_{CS}~,\\
\mathcal{A}_{CS}^{(c)} &= -i g\sqrt{Z} {4m_R^2} C(\eta_p) e^{i\sigma_0}  \left(\frac{\sqrt{r_0 r_0^{C}}}{8\pi}\right) T_{CS}~,
\end{align}
with the
complex-valued integral
\begin{align}
\mathcal{I} = \int \frac{d^3\mathbf{q}}{(2\pi)^3} \frac{C^2(\eta_q) e^{2\eta_q 
 \arctan (| \mathbf{q} | /\gamma_0)}} {\mathbf{q}^2+\gamma_0^2}  
\frac{1}{\mathbf{p}^2-\mathbf{q}^2 + i\epsilon}~.
\end{align}
The total amplitude $\mathcal{A}$ is the sum of the amplitudes with and
without resonance $\mathcal{A} =\mathcal{A}_{C}^{(a)} + \mathcal{A}_{CS}^{(b)} + \mathcal{A}_{CS}^{(c)}$.

At LO, the Coulomb-modified effective range in the
$^{10}\text{Be}-p$ system is zero and the amplitude reduces to
\begin{equation}
  \label{eq:tcs-neg_a}
T_{CS} = -\frac{2\pi }{m_R} \left[ \frac{1}{ -1/a_0^C - 2k_C H(\eta_p)}   
\right]~.
\end{equation}

\paragraph{\bf Results without resonant final state interactions.}
We consider two scenarios: beta-delayed proton emission with and
without resonant final state interactions from a low-lying resonance in $^{11}$B.
We start with the first scenario and
use the one-neutron separation energy of $^{11}$Be
$S_n=0.5016$~MeV~\cite{Kelley:2012qua}. 
In Fig.~\ref{fig:gamma-be11},
we plot the differential decay rate $d\Gamma/dE$ as a function of the
kinetic energy $E$ of the outgoing hadrons. The solid line gives the
result obtained by Baye and Tursunov~\cite{Baye:2010cj}.
The dash-dotted line shows the EFT result with an uncertainty band
obtained by adding an uncertainty of order 
$R_{\text{core}}/R_{\text{halo}}\approx
40$~\% from higher order corrections where we use the smallest value of $R_{\text{halo}}$ given by $1/\gamma_0$ while we estimate $R_{\text{core}}$ by the effective range $r_0$ as a conservative estimate.  
The remaining curve includes resonant
final state interactions and will be discussed below.

For the branching ratio, we obtain
$b_p=\Gamma/\Gamma_{\rm total} = (1.31 \pm 0.51) \times 10^{-8}$ 
where the EFT uncertainty is again estimated to be of the order of 40~\%.
Correspondingly, we obtain for the decay rate
$\Gamma = (6.6 \pm 2.6)\times 10^{-10}~{\rm s}^{-1}$.  Baye and 
Tursunov~\cite{Baye:2010cj}
obtain $\Gamma=1.5\times 10^{-9}~{\rm s}^{-1}$ which differs by a 
factor of $2.3$ from our result. We note, however, that they 
used a Woods-Saxon potential with Coulomb interactions tuned to reproduce 
$^{11}$B properties in the final state.  Both theoretical results are 
significantly smaller than the experimental results reported in 
Refs.~\cite{Borge:2012nz,Riisager:2014rma,Riisager:2014gia,Ayyad:2019kna}.

\paragraph{\bf Results with resonant final state interactions.}
\begin{figure}[t]
  \centering
  \includegraphics[width=0.45\textwidth]{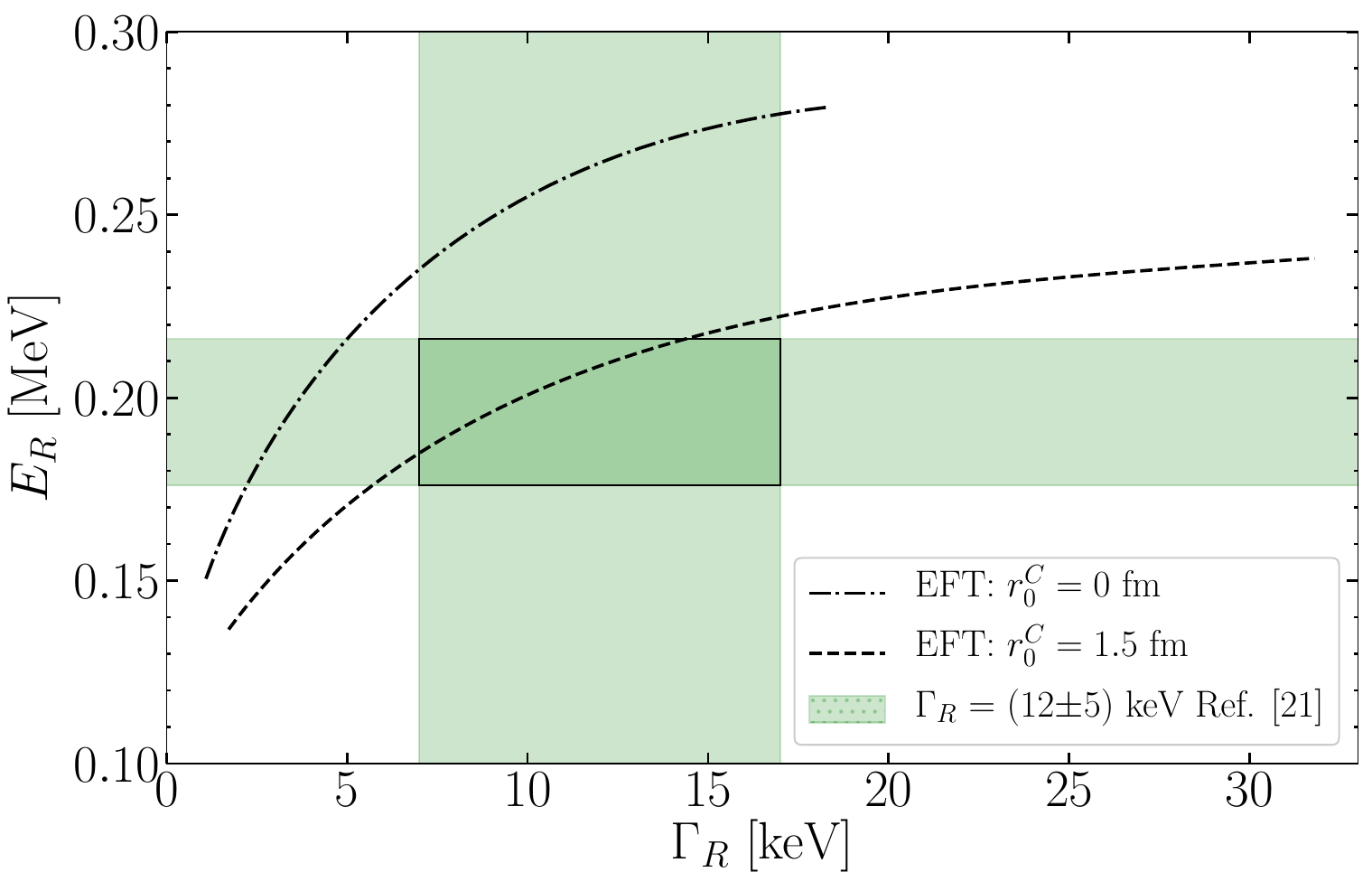}
  \caption{Possible resonance parameter combinations fulfilling the sum rule.
		The dash-dotted line shows the combinations for $r_0=0$~fm at LO 
		corresponding to $r_0^C=0$~fm while the dashed line shows the combinations
		for $r_0=2.7$~fm at NLO corresponding to $r_0^C=1.5$~fm. The green bands show
		the resonance parameters given in Ref.~\cite{Ayyad:2019kna}.
		}
  \label{fig:resonance-params}
\end{figure}
We now discuss the second scenario including final state
interactions.
In Fig.~\ref{fig:resonance-params}, we show the possible resonance parameter 
combinations that fulfill the beta-strength sum rule. The dash-dotted line
is the result at LO where the effective range in the incoming channel as
well as the Coulomb-modified effective range in the outgoing channel are zero.
At NLO, we use $r_0=2.7$~fm determined in Ref.~\cite{Hammer:2011ye} from 
the measured $B$(E1) strength for Coulomb dissociation of $^{11}$Be. The 
one-neutron separation energy as well as the effective range of $^{11}$Be 
determine the Coulomb-modified effective range in the outgoing channel to be 
$r_0^C=1.5$~fm. The sum rule is then satisfied to very good approximation for 
a wide range of Coulomb-modified scattering lengths in the outgoing channel.
The square shows the experimentally measured resonance parameter combinations 
given in Ref.~\cite{Ayyad:2019kna}. 
We note that the value of $r_0^C$ is determined independently from the experimental resonance parameters. 
Our NLO curve depicted as the dashed line corresponding to $r_0^C=1.5$~fm
exhibits combinations of $E_R$ and $\Gamma_R$ that are in agreement with this measurement as 
indicated by the overlap of the square and the curve.

\begin{figure}[t]
  \includegraphics[width=0.45\textwidth]{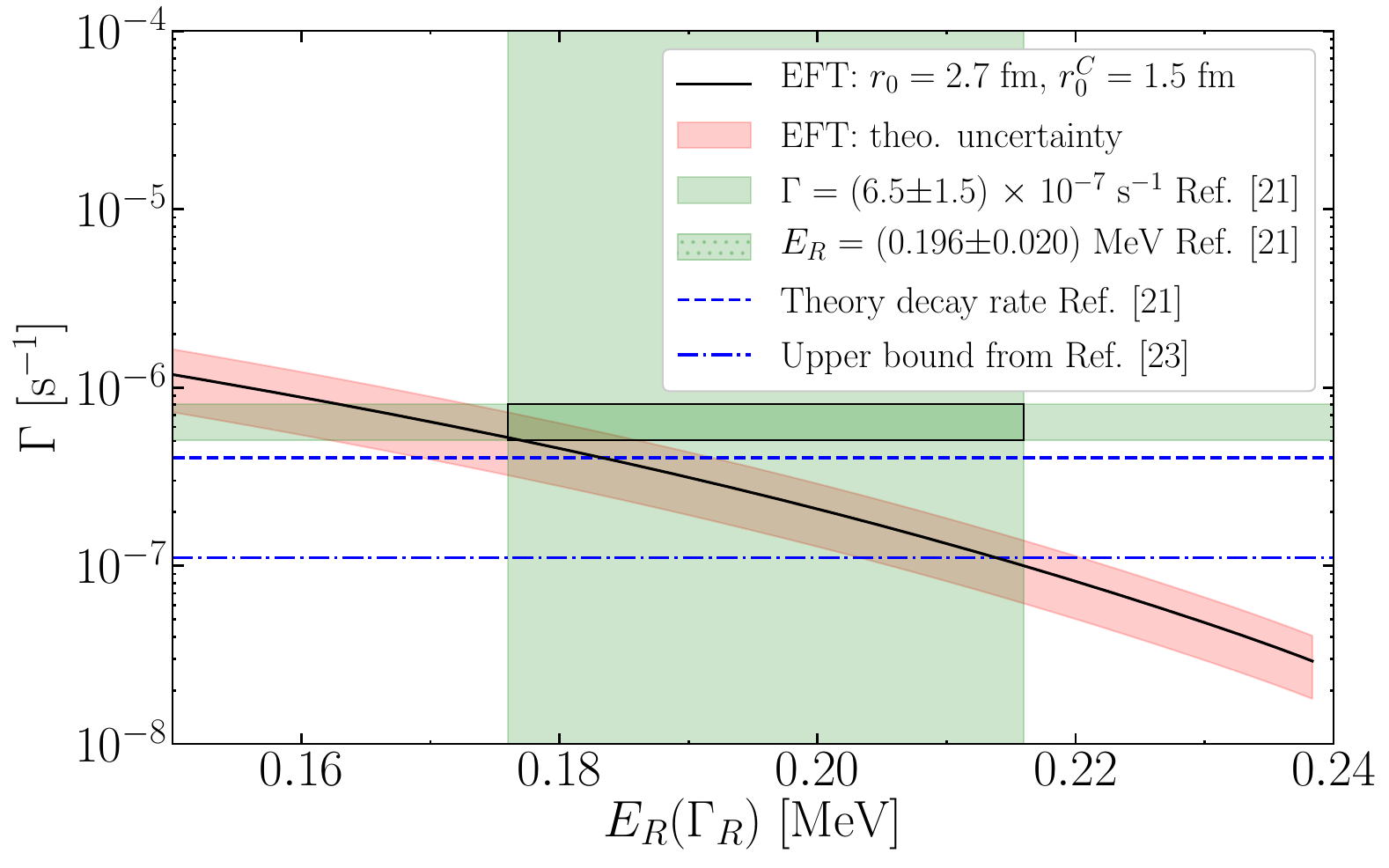}
    \caption{Partial decay rate as a function of the resonance energy
			at NLO using the corresponding resonance width in accordance with
			the sum rule (see Fig~\ref{fig:resonance-params}). Explanation of 
			curves and bands is given in inset.
			}
    \label{fig:resonance}
\end{figure}

In Fig.~\ref{fig:resonance}, we show the results for the decay rate as
a function of the resonance energy at NLO while using the
corresponding resonance width that satisfies the sum rule as shown in
Fig.~\ref{fig:resonance-params}.  The black line represents the decay
rate obtained moving along the NLO curve in
Fig.~\ref{fig:resonance-params} while the red shaded envelope gives
the theoretical uncertainty estimated from the counterterm 
contribution in the axial current scaling with 
$R_{\text{core}}/R_{\text{halo}} \approx 40$~\%. The green bands show the
experimentally measured branching ratio and resonance energy of
Ref.~\cite{Ayyad:2019kna}.  The horizontal blue dashed line denotes
the result of the model calculation carried out in
Ref.~\cite{Ayyad:2019kna} whereas the horizontal blue dash-dotted line
gives the upper bound of Ref.~\cite{Riisager:2020glj}.  Comparing our
results with Ref.~\cite{Riisager:2020glj}, we find that resonance
energies $E_R\geq 0.214$ MeV give results compatible with this upper
bound. The corresponding resonance widths can be read off in
Fig.~\ref{fig:resonance-params}.  When comparing our results with
Ref.~\cite{Ayyad:2019kna}, we find that the low-lying resonance
measured in Ref.~\cite{Ayyad:2019kna} with $E_{R}=0.196(20)$ MeV and
width $\Gamma_R = 12 (5)$ keV is consistent with their experimentally
measured branching ratio as indicated by the overlap of the square and
the red shaded band.  According to Fig.~\ref{fig:resonance-params}, we
determine the width corresponding to the resonance energy
$E_R=0.196(20)$ MeV as $\Gamma_R = (9.0^{+4.8}_{-3.3}\text{(exp.)}^{+5.3}_{-2.2}\text{(theo.)})$~keV, which
agrees well with the experimental value.  
At LO, the resonance width scales as $k_C^2/m_R$ whereas at NLO this value is enhanced by a factor of $1/(1-3k_Cr_0^C)$. This enhancement for Coulomb halos is well known \cite{Ryberg:2015lea,Luna:2019ufu,Schmickler:2019ewl}.
Using $E_{R}=0.196(20)$ MeV,
we calculate the logarithm of the comparative half-life
log$(ft)=3.0$ with $B_{\text{GT}}=2.88$ and $B_{\text{F}}=0.96$ for a
decay including both Gamow-Teller and Fermi transitions and
log$(ft)=3.1$ with $B_{\text{GT}}=2.88$ for a pure Gamow-Teller
transition. The latter result can be compared to log$(ft)=4.8(4)$
calculated by Ayyad {\it et al.}~\cite{Ayyad:2019kna} which was obtained
using a pure Gamow-Teller transition as well, but is significantly
larger than our result. This large log$(ft)$ value was also criticized
in the comment by Fynbo {\it et al.}~\cite{Fynbo:2019dfo}.  
Ayyad {\it et al.} corrected the value to log$(ft)=2.8(4)$ in their recent
erratum \cite{Ayyad:2019kna}. This new value is now in 
good agreement with our result.
Using the half-life for $^{11}$Be given in Ref. \cite{Kelley:2012qua} we convert
the Halo EFT result for $E_{R}=0.196(20)$~MeV and
$\Gamma_R = (9.0^{+4.8}_{-3.3}\text{(exp.)}^{+5.3}_{-2.2}\text{(theo.)})$~keV into the final result for the
branching ratio
$b_p = 4.9_{-2.9}^{+5.6}\text{(exp.)}_{-0.8}^{+4.0}\text{(theo.)}
  \times 10^{-6}$.  The corresponding differential decay rate is shown
by the dashed line in Fig.~\ref{fig:gamma-be11}.

\paragraph{\bf Conclusion.}
In this paper, we considered $\beta$-delayed proton emission from
$^{11}$Be. We compared the scenario with no strong final state
interactions with the scenario of a resonant enhancement in the final
$^{10}$Be$-p$ channel up to NLO. In the case of no strong final state
interactions, we obtained results that are in qualitative agreement
with Baye and Tursunov with remaining small differences that can be
explained by the different treatment of the final state
channel. Including a low-lying resonance with the energy measured in
Ref.~\cite{Ayyad:2019kna} results in a resonance width and partial
decay rate in agreement with this experiment. Thus, our
model-independent calculation supports the experimental finding of a
low-lying resonance.\footnote{See Ref.~\cite{Okolowicz:2019ifb} for
  another recent theoretical calculation in support of this
  resonance.}
Furthermore, we have 
explored the sensitivity of the partial decay
rate to the resonance energy and decay width and found that this
problem is fine tuned, {\it i.e.} only certain combinations of width
and resonance energy can reproduce the partial decay rate.
In contrast to  the model calculation in Ref.~\cite{Ayyad:2019kna}, we
included both, Fermi and Gamow-Teller transitions.
However, if a pure Gamow-Teller decay is considered, their partial decay rate can also be reproduced with slightly smaller resonance parameters.
Thus, our result implies that $^{11}$Be is not a good laboratory to detect dark neutron decays since no exotic mechanism is needed to explain the partial decay rate.

The uncertainties are largely determined by higher order contributions
of the EFT expansion. The next contribution within our power counting
that we did not include is a counterterm contribution in the axial
current scaling with $R_{\text{core}}/R_{\text{halo}}$.
Uncertainties of the $S$-wave input parameter (the one-neutron
separation energy) do not impact the total uncertainty
significantly. Therefore, we estimate the uncertainty in the final
decay rate to be approximately
$R_{\text{core}}/R_{\text{halo}} \approx 40$~\%.  Experimental
data with higher precision could be used to constrain the
$^{10}$Be$-n$ and $^{10}$Be$-p$ interactions.  It will be interesting
to test whether the inclusion of this resonance changes the Halo EFT
predictions for deuteron induced neutron transfer reactions off
$^{11}$Be which were investigated in Ref.~\cite{Schmidt:2018doj}.

\begin{acknowledgments}
  We acknowledge useful discussions with
  T.~Papenbrock, M.~Madurga and  thank D.~Baye for providing the
  data published in Ref.~\cite{Baye:2010cj}.
  WE thanks the Nuclear Theory groups of UT Knoxville and Oak Ridge
  National Laboratory for their kind hospitality and support during
  his stay.  
  HWH and LP thank the Institute for Nuclear Theory at the University of
  Washington for its kind hospitality and stimulating research environment.
  This research was supported in part by the INT's U.S. Department of
  Energy grant No. DE-FG02- 00ER41132. It has been funded
  by the Deut\-sche For\-schungs\-ge\-mein\-schaft (DFG, German
  Research Foundation)~-- Project-ID 279384907~-- SFB 1245,
  by the German Federal Ministry of Education and Research (BMBF)
  (Grant no. 05P18RDFN1),
  by the National Science Foundation under Grant No.  PHY-1555030, and
  by the Office of Nuclear Physics, U.S.  Department of Energy under
  Contract No. DE-AC05-00OR22725.
\end{acknowledgments}

\end{document}